\documentclass{JHEP3}
\usepackage{amsmath,amssymb}
\usepackage{graphicx,epsfig}

\newcommand{\Fbar}[1]{\hspace{0.3em}{\overline{\rule[0.75em]{1.15em}{0em}}\hspace{-1.3em}F{#1}}}
\newcommand{\no}{\nonumber \\}

\newcommand{\non}{\nonumber}
\newcommand{\fs}{\; .}
\newcommand{\co}{\; ,}  
\newcommand{\strichpunkt}{|}
\newcommand{\eps}{0}
\newcommand{\epsj}{\epsilon}
\newcommand{\quadp}{\hspace{1.3em}}

\newcommand{\mubar}{\bar{\mu}}
\newcommand{\Fdiv}{F_{\rm div}}

\preprint{CPT-P06-2006\\
UWThPh-2006-3\\
hep-ph/0603153\\
May 2006}

\title{The expansion by regions in $\pi K$ scattering}

\author{Roland Kaiser\footnote{Corresponding author}\\
Centre de Physique Th\'{e}orique,\footnote{Unit\'{e} mixte de
recherche (UMR 6207) du CNRS et des Universit\'{e}s Aix-Marseille I,
Aix-Marseille II, et du Sud Toulon-Var; laboratoire affili\'{e} \`{a}
la FRUMAM (FR 2291).}  \hspace{0.1em}CNRS-Luminy,\\ 
Case 907, F-13288 Marseille Cedex 9, FRANCE\\
E-mail:  \email{kaiser@cpt.univ-mrs.fr}
}
\author{Julia Schweizer\\
Institute for Theoretical Physics, University of Vienna,\\ 
Boltzmanngasse 5, A-1090 Vienna, AUSTRIA}

\abstract{
We discuss a number of two loop (vertex) integrals relevant for $\pi K$ scattering at threshold. As such these are functions of two well separated mass scales, $M_\pi/M_K \ll 1$. The method of regions allows an expansion in this mass ratio prior to the integration. In the cases considered, the coefficients in this expansion can be evaluated with elementary methods. {As a first example of an application we present the leading strange quark contributions to the pion mass.}
}


\begin{document}

\section{\label{sec:level1} Introduction}
Two loop calculations are not a rarity anymore in chiral perturbation theory \cite{Schenk:1993ru,Bellucci:1994eb,
Golowich:1995kd,Burgi:1996qi,Bijnens:1995yn,Post:1997dk,Toublan:1997rr,Bijnens:1998fm,Bijnens:1998yu,Bijnens:1996wm,GK,Gasser:1998qt,Amoros:1999dp,Durr:1999dp,Amoros:2000mc,Post:2000gk,Bijnens:2004eu,Bijnens:2004bu,Bijnens:2004hk,Gasser:2005ud,
Bijnens:2006jv,Colangelo:2006mp,Bijnens:2006ve}. Due to the presence of masses the calculations are rather complex and only few explicit analytic results are available, particularly in the three flavour case.  

In cases where the final result depends on well separated mass scales it is reasonable to expect that a few terms in the expansion in terms of the ratios of the scales yield a good approximation to the full result. In turn, there is hope that the individual terms in the expansion represent simpler expressions and may even be given in closed form. Because of the complexity of the full results, it is desirable to have a method which allows an expansion prior to the integration. Moreover, integral representations suitable for numerical integration for generic momenta sometimes exhibit singularities at the (pseudo) thresholds.

Below, we discuss integrals that contribute to the $\pi K$ scattering amplitude at threshold and thus depend only on the pion and kaon masses $M_\pi$ and $M_K$ (throughout, we work in the isospin limit $m_u = m_d$). 
The envisaged physical application is to find analytic representations for the expansion of the $\pi K$ scattering lengths in $M_\pi/M_K $. The most precise values for the $\pi K$ scattering lengths are presently obtained from an analysis of Roy-Steiner equations \cite{Ananthanarayan:2000cp,Buettiker:2003pp}. Alternatively, particular combinations of $\pi K$ scattering lengths may be extracted from experiments on $\pi K$ atoms \cite{Deser:1954vq,Bilenky:1969zd,Schweizer}. The $\pi K$ atom decays due to the strong interactions into $\pi^0 K^0$ and a lifetime measurement will allow one to determine the isospin odd S-wave $\pi K$ scattering length $a_0^-=\frac{1}{3}(a^{1/2}_0-a^{3/2}_0)$. Such a measurement is planned at CERN \cite{futureDIRAC}. 
The scattering length $a_0^-$ differs from other low-energy parameters in $\pi K$ scattering 
in that it is protected against contributions of $M_K^2$ in the chiral expansion:
In the framework of SU(2) chiral perturbation theory \cite{Weinberg:1978kz,Gasser:1983yg,Roessl:1999iu}, where the kaon is treated as heavy, there exists a low-energy theorem \cite{Roessl:1999iu} which states that the Weinberg current algebra result \cite{Weinberg:1966kf,griffith} receives corrections of order $M_\pi^2$ only,
 \begin{equation}
   a^-_0 = \frac{M_\pi M_K }{8\pi
   F_\pi^2(M_\pi+M_K)}\left\{1+O(M_\pi^2)\right\}.
   \label{eq: Roessl}
 \end{equation}
  Here $M_\pi$, $M_K$ and $F_\pi$ denote the physical meson masses and the physical pion decay constant. 
It was therefore expected that the one-loop result \cite{Bernard:1990kw,Kubis:2001bx,Nehme:2001wa} in the SU(3) theory \cite{Gasser:1984gg} 
should represent a decent estimate for this scattering length. However, the dispersive analysis from Roy-Steiner equations \cite{Buettiker:2003pp} and the chiral two-loop calculation \cite{Bijnens:2004bu} are not in agreement with this expectation. In fact, the numerical analysis performed in Ref. \cite{Bijnens:2004bu} showed that the two-loop order corrections to $a_0^-$ are of the same order of magnitude as the one-loop contributions. 
To understand the nature of these rather substantial corrections, a partial analysis of $a_0^-$ at next-to-next-to-leading order in SU(3) chiral perturbation was performed in Ref. \cite{Schweizer:2005nn}. 
The present work represents a first step towards an analytic two-loop representation for the expansion of the $\pi K$ scattering length in $M_\pi/M_K $.

The technology we employ to obtain our results is known by the name of expansion by regions 
\cite{Beneke:1997zp,Smirnov:2002pj}. 
The recipe for the method is simple: i) In the integrand identify all the relevant integration regions, ii) expand the integrand in each region, iii) sum up the integrals (integrated over the full integration domain) of all the contributions. The method obviously owes its applicability to the fundamental property of dimensional regularization, 
being that the integral only receives contributions from scales which are present in the integrand. 
The divergences encountered in intermediate steps are also tamed by dimensional regularization.  
The number and nature of the regions to be considered depends on the problem at hand. In the present case, it is sufficient to consider hard ($ M_K$) and soft ($  M_\pi$) regions. Since we furthermore do not violate manifest Lorentz invariance we only need to consider two distinct regions for the loop momenta. 

In Section~\ref{sec:J-integral}, we introduce the method for the example of the two point integral $J$ which contributes to $\pi K$ scattering at the one loop level. On the basis of the comparison with the expansion of the known full result we are able to demonstrate the validity of the method in this case.
The prescription translates to the case of two loop integrals which are considered in Section~\ref{sec:fish}. Specifically, we apply the method to three variants of fish type (two loop vertex) integrals. We demonstrate the usefulness of the method by computing the two leading terms in the expansion of $m/M $ of each integral. 
Apart from products of one loop integrals (see Appendix~\ref{app: Table of one loop Integrals}), the determination of the coefficients in the expansion involves several variants of on-shell sunset type integrals whose evaluation is described in Appendix~\ref{app: Two loop Integrals}.   
Finally, Section~\ref{sec:conclusions} contains our conclusions. 

In the present work, we are making use of methods that are common in fields other than chiral perturbation theory. While we are unable to give a comprehensive bibliography on the subject here, this is mitigated by the existence of excellent reviews \cite{Smirnov:2002pj,Smirnov:2004ym}. 
Ref.~\cite{Beneke:1997zp} is equally suited for an introduction to the subject of the expansion by regions. Note that while at present there are no proofs for the validity of the method in the Minkowskian region, there are also no known counterexamples.
The expansion in a small mass ratio $m/M$ was studied in Ref.~\cite{Czarnecki:1996nr} in the case of the two loop two point function and in the framework of nonrelativistic QED in Refs.~\cite{Czarnecki:2000fv,Blokland:2001fn}, with techniques very similar to ours. For two loop integrals with arbitrary masses, see Refs.~\cite{Davydychev:1992mt,Ghinculov:1994sd,Post:1996gg,Ghinculov:1997pd}.   
Last but not least let us mention the existence of other methods suitable for the expansion of Feynman integrals in small parameters. A prominent example is the Mellin-Barnes representation \cite{BjorkenWu,TruemanYao,Polkinghorne,Greub:1996tg,Greub:2000sy,Asatryan:2001zw,Bieri:2003ue,Friot:2005cu,Smirnov:2004ym}. 
\section{$J$ Integral}
\label{sec:J-integral}
 
\begin{figure}[t]
\begin{center}
\epsfig{figure=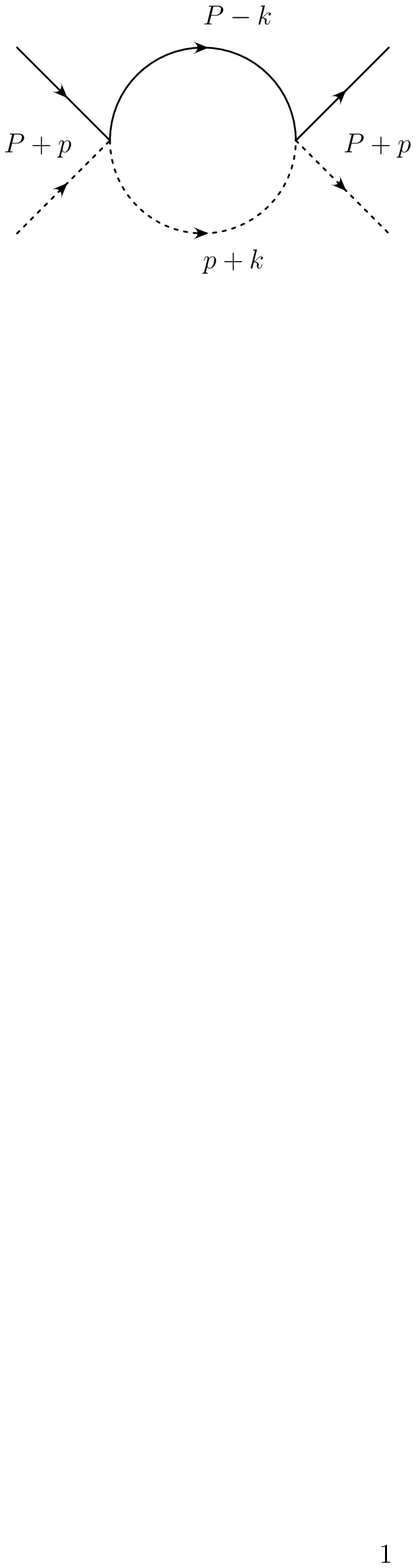,height=4cm,bbllx=120,bblly=645,bburx=300,bbury=778,clip}
\caption{$J$ Integral: The solid line denotes the kaon, the dashed line stands for the pion.}
\label{fig: JIntegral}
\end{center}
\end{figure}

 As an illustration of the method, we consider the $J$ Integral at threshold\footnote{An analogous consideration for the on-shell integral, $(p+P)^2=M^2$, may be found in Ref.~\cite{Smirnov:2002pj}.}
 $(p+P)^2=(m+M)^2$,
\begin{align}
\label{Jthr}
J_{\rm thr}(m,M) & =\frac{1}{i}\int \frac{d^dk}{(2\pi)^d}\frac{1}{-2p k-k^2-i\eps}\frac{1}{2Pk-k^2-i\eps} 
\\
 & = J_{\rm thr}(0,M)  - \frac{1}{(4\pi)^2} \frac{m}{m+M} \ln \frac{m^2}{M^2} + O(d-4) 
 \fs \non
\end{align}
In order to simplify our notation, we use throughout the symbols $ M=M_K $ and $m=M_\pi $.
As shown in Fig.~\ref{fig: JIntegral}, we let the external kaon momentum  $P^2=M^2$ flow through the kaon line so that the kaon mass is cancelled. Otherwise the momentum assignment is free. We choose a symmetric configuration where  $p^2=m^2$ denotes the external pion momentum.
Further our reference frame is chosen such that $P\sim M$ and $p \sim m$ (center of mass system). 
The regions for the loop momenta $k$ are defined as follows
\begin{alignat}{2}
&\textrm{hard (h)}: & \quad \quad & k \sim M
\co \non\\
&\textrm{soft (s)}: & \quad  \quad & k \sim m
\fs
\end{alignat}

The expansion of the hard region ($k \sim M$) looks as follows:
\begin{align}
J^{\rm h}(m,M) &=\frac{1}{i}\int \frac{d^dk}{(2\pi)^d}\frac{1}{-k^2-i\eps}\frac{1}{2Pk-k^2-i\eps} \sum^\infty_{n=0}\left( \frac{2pk}{-k^2-i\eps}\right)^n
\non\\
&=\sum_{n=0}^\infty J^{\rm h}_n
\fs
\end{align}
Taking the measure into account, the resulting order, $J^{\rm h}_n  = O(m^n/M^n)$, is anticipated. Studying the behaviour at zero and infinity shows that the leading order term $J^{\rm h}_0$, coinciding with $J_{\rm thr}(0,M)$, contains an ultraviolet divergence and is infrared finite. Explicitly, we get 
\begin{align}
J^{\rm h}_0
&=J_{\rm thr}(0,M)  = \frac{ \mubar^{-2\epsj}}{(4\pi)^2}\left(\frac{1}{\epsj_{\rm UV}}-\ln\frac{M^2}{\mu^2}+2 \right)+O\left(\epsj\right) 
\fs
\end{align}
We indicate the nature of the divergence by $\epsj\rightarrow\epsj_{\rm UV}$, where 
\begin{align}
\epsj = 2-\frac{d}{2} \co  \quad \mubar^{2} = \frac{e^{-\Gamma'(1)}}{4\pi} \, \mu^{2} \fs 
\label{eq: epsj}
\end{align}
For $n \ge1$, the hard part is ultraviolet finite and infrared divergent. The next-to-leading order takes the form
\begin{align}
J^{\rm h}_1
&=-\frac{m}{M}\frac{\mubar^{-2\epsj}}{(4\pi)^2}\left(\frac{1}{\epsj_{\rm IR}}-\ln\frac{M^2}{\mu^2}+2\right)+O\left(\epsj\right)
\co 
\end{align}
where $\epsj\rightarrow \epsj_{\rm IR}$. 

The expansion of the soft region ($k \sim m$) takes the form
\begin{align}
J^{\rm s}(m,M) &=\frac{1}{i}\int \frac{d^dk}{(2\pi)^d}\frac{1}{-2pk-k^2-i\eps}\frac{1}{2Pk-i\eps}\sum^\infty_{n=1}\left( \frac{k^2}{2Pk-i\eps}\right)^{n-1} 
\no
&=\sum_{n=1}^\infty J^{\rm s}_n
\fs
\end{align}
Again, $J^{\rm s}_n  = O(m^n/M^n)$. In this case there is an ultraviolet divergence for all $n(>0)$ and the integrals are infrared finite. For the leading term in $m/M$ we find
\begin{align}
J^{\rm s}_1 &= 
\frac{m}{M} J_{\rm thr}(m,0)
=\frac{m}{M} \frac{\mubar^{-2\epsj}}{(4\pi)^2}\left(\frac{1}{\epsj_{\rm UV}}-\ln\frac{m^2}{\mu^2}+2 \right)+O\left(\epsj\right).
\end{align}
This result is obtained easily by choosing the frame such that $ P = M/m\, p $ and subsequent partial fractioning.    
In the sum of the hard and soft part at order $m/M$
\begin{align}
J^{\rm h}_1 +J^{\rm s}_1 =- \frac{1}{(4\pi)^2}\frac{m}{M}\ln \frac{m^2}{M^2}
\co
\end{align}
the infrared pole term in $J^{\rm h}_1$ and the ultraviolet pole term in $J^{\rm s}_1$ have cancelled (since $\epsj_{\rm UV}=\epsj_{\rm IR} =\epsj$) to give the desired result as in Eq.~(\ref{Jthr}).  
The cancellation of an infrared with an ultraviolet divergence is in fact common in dimensional regularization, as the following example shows~\cite{Manohar:1997qy}:
\begin{align}
\int_k 
\frac{1}{(-k^2-i\eps)^2}&=
\int_k
\left\{  \frac{1}{(-k^2-i\eps)(m^2-k^2-i\eps)}+\frac{m^2}{(-k^2-i\eps)^2(m^2-k^2-i\eps)}\right\} \no
&= \frac{1}{(4\pi)^2}\left( \frac{1}{\epsj_{\rm UV}}-\frac{1}{\epsj_{\rm IR}}\right)  = 0 
\fs
\end{align}
In the following, we will refrain from detailing the nature of the divergences encountered at intermediate steps and trust in dimensional regularization 
to take care of the necessary cancellations.
This procedure works on the premise that the divergence structure of the end result is known from other sources. This is of course the case for the $J$ integral as well as for the two loop diagrams considered in the next section.

To conclude the present section we note that it is straightforward to establish the relations 
\begin{align}
J_{n+1}^h = -\frac{m}{M} J_n^h \co \quad J^s_{n+1} = -\frac{m}{M} J_n^s  
\co
\end{align}
hereby extending the validity of the approach in the case of the integral $J_{\rm thr}(m,M)$ to all orders in $m/M$. The method is equally valid when the dimension $d$ is left arbitrary.    


\section{Fish type integrals}
\label{sec:fish}
In this section, we discuss the two loop fish type integrals occurring in $\pi K$ scattering at threshold, $(p+P)^2=(m+M)^2$.
Our aim is an expansion of the fish type diagrams in powers of the mass ratio $m/M$. To achieve this we perform an expansion by regions. 
The set of regions that contribute to the expansion of the fish type diagrams at threshold in general consists of 5 regions~\cite{Blokland:2001fn},
\begin{alignat}{5}
&\textrm{h-h} &&:  \quad\quad & k & \sim M\co \quad  & l & \sim M \co \quad & k-l & \sim M
\co \no
&\textrm{h-h'} &&:  \quad\quad & k & \sim M\co \quad & l & \sim M \co \quad & k-l & \sim m 
\co \no
&\textrm{s-h} &&: \quad\quad & k & \sim m\co \quad   &l &\sim M \co &&
\no
&\textrm{h-s} &&: \quad\quad & k & \sim M\co\quad    &l &\sim m \co &&
\no
&\textrm{s-s} &&: \quad\quad & k & \sim m\co \quad   & l &\sim m \fs &&
\end{alignat}
As indicated, the difference in the h-h and h-h' regions lies in the fact that the difference $k-l$ is counted as hard and soft, respectively. Note that in the examples considered below, the h-h' region does not contribute. The loop momenta $k$ and $l$ are shown for example in Fig.~\ref{fig: pipiloop}. In the following, we discuss this topology with different particles running in the subgraph.


\subsection{Pion loop}
\begin{figure}[t]
\begin{center}
\epsfig{figure=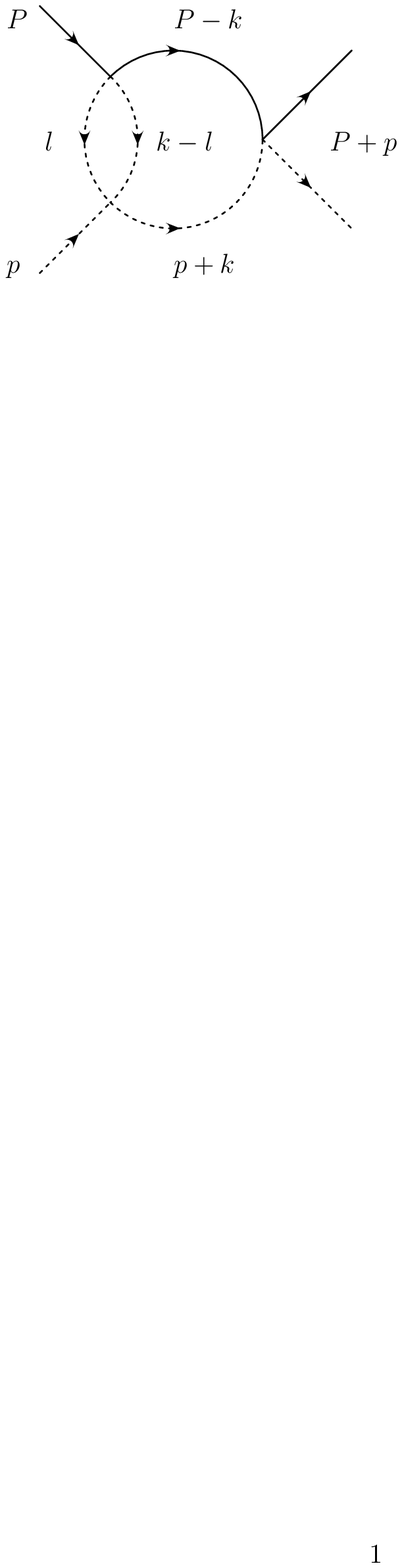,height=4cm,bbllx=120,bblly=645,bburx=300,bbury=778,clip}
\caption{Fish diagram with an internal pion loop.}
\label{fig: pipiloop}
\end{center}
\end{figure}
Let us start with the fish diagram containing an internal pion loop (for our notation, see Appendix \ref{app: Two loop Integrals})
\begin{align}
F1 = \int_{k,l}  \frac{1}{m^2-l^2-i\eps}\frac{1}{m^2-(k-l)^2-i\eps}\frac{1}{2Pk-k^2-i\eps}\frac{1}{-2pk-k^2-i\eps} 
\co
\end{align}
shown in Fig. \ref{fig: pipiloop}. The fish type diagram contains a nonlocal divergence, which is known for arbitrary masses and external momenta, see Eq.~(\ref{eq: fish div}). At threshold no additional infrared singularities occur.

The h-h region contributes at order one, the s-s part is of order $m/M$, while the $\textrm{h-h'}$ and $\textrm{h-s}$ regions start contributing at order $m^2/M^2$ which is beyond the scope of our calculation. The expansion in the s-h
region leads to scaleless integrals so that this part vanishes altogether. We thus have
\begin{align}
F1 = F1^{\rm hh}_0+F1^{\rm hh}_1+F1^{\rm ss}_1+O\left(\frac{m^2}{M^2}\right) 
\fs
\end{align}
The expansion of the h-h region ($k\sim M$, $l \sim M$) looks as follows
\begin{align}
F1^{\rm hh} &= \int_{k,l}\frac{1}{-l^2-i\eps}\frac{1}{-(k-l)^2-i\eps}\frac{1}{2Pk-k^2-i\eps}\frac{1}{-k^2-i\eps}\left\{ 1+\frac{2p k}{-k^2-i\eps}+O\left(\frac{m^2}{M^2}\right)\right\} \no
&=F1^{\rm hh}_0+F1^{\rm hh}_1+O\left(\frac{m^2}{M^2}\right)
\fs
\end{align}
To evaluate the h-h contribution at leading order, we first carry out the integration over the subgraph momentum $l$. The resulting integral over $k$ is then again of the form of a one loop integral and the result can be written in terms of one loop functions,
\begin{align}
F1^{\rm hh}_0 = M^{2d-8}I_{1,1}\{0,0;-1\}I_{3-\frac{d}{2},1}\{0,1;1\}   
\co
\end{align}
where the $I_{\alpha,\beta}$ are defined in Appendix~\ref{app: Table of one loop Integrals}. At leading order in $m/M$ the fish type integral $F1$ is given by
\begin{align}
F1_0 &=F1^{\rm hh}_0
\\ 
& = \frac{\mubar^{-4\epsj}}{(4\pi)^4}\left\{\frac{1}{2\epsj^2}+\frac{1}{2\epsj}\left[5-2\ln \frac{M^2}{\mu^2}\right]+\ln^2 \frac{M^2}{\mu^2} -5\ln \frac{M^2}{\mu^2}+\frac{19}{2}+\frac{5\pi^2}{12}\right\} +O(\epsj) 
\co \non  
\end{align}
where the parameters $\epsj$ and $\mubar$ are defined in Eq.~(\ref{eq: epsj}). For the next-to-leading order contribution,  we find
\begin{align}
F1^{\rm hh}_1 = -\frac{m}{M}\, F1^{\rm hh}_0 \fs 
\end{align}

To evaluate the contribution from the s-s region ($k\sim m$, $l\sim m$) at order $m/M$
\begin{align}
F1^{\rm ss}_1 = \int_{k,l}  \frac{1}{m^2-l^2-i\eps}\frac{1}{m^2-(k-l)^2-i\eps}\frac{1}{2Pk-i\eps}\frac{1}{-2pk-k^2-i\eps} 
\co 
\end{align}
we exploit the fact that our reference frame can be chosen such that $p=m/M P$. 
Now we may use the following trick \cite{'tHooft:1978xw,Chetyrkin:1981qh,Vassiliev}: In the integral we insert the identity
\begin{align}
1 = \frac{1}{d} \frac{\partial }{\partial k_\mu}k_\mu
\end{align}
and integrate by parts 
to obtain a relation between the s-s part and 
sunset type diagrams 
\begin{align}
F1^{\rm ss}_1 = 
\frac{m}{M} \frac{m^{2d-8}}{d-4} \left[2 S_{1,1,2}\{1,1,1;1\}+S_{2,1\strichpunkt 1}\{1,1;0,1\} -I_2\{1\} I_{1\strichpunkt1}\{1;1,1\}\right] 
\fs
\end{align}
The one loop integrals $I_\alpha $ and $I_{\alpha|\beta}$ are listed in Appendix~\ref{app: Table of one loop Integrals}. The sunset $S_{2,1\strichpunkt 1}\{1,1;0,1\}$, given in Appendix~\ref{app: Two loop Integrals}, can be expressed in terms of Gamma functions. To evaluate the finite part of $F1^{\rm ss}_1$ we further need the 
equal mass case on-shell sunset $S_{1,1,2}\{1,1,1;1\}$ at order $\epsj$. The evaluation of this sunset type diagram is outlined in Appendix~\ref{app: Two loop Integrals} and the result at order $\epsj$ is listed in Eq.~(\ref{eq: sunset}).

If we add the contributions from the h-h and s-s regions, the result for the fish diagram $F1$ at order $m/M$ yields
\begin{align}
F1_1&= F1_1^{\rm hh} + F1_1^{\rm ss} 
\no
&= \frac{m}{M}\frac{ \mubar^{-4\epsj}}{(4\pi)^4}\left\{  -\left[ \frac{1}{\epsj} - 2 \ln \frac{M^2}{\mu^2} \right] \ln \frac{m^2}{M^2} + \ln^2 \frac{m^2}{M^2}-5\ln \frac{m^2}{M^2}-3 \pi^2\right\} +O(\epsj) \fs
\end{align}

\subsection{Kaon loop}
\begin{figure}[t]
\begin{center}
\epsfig{figure=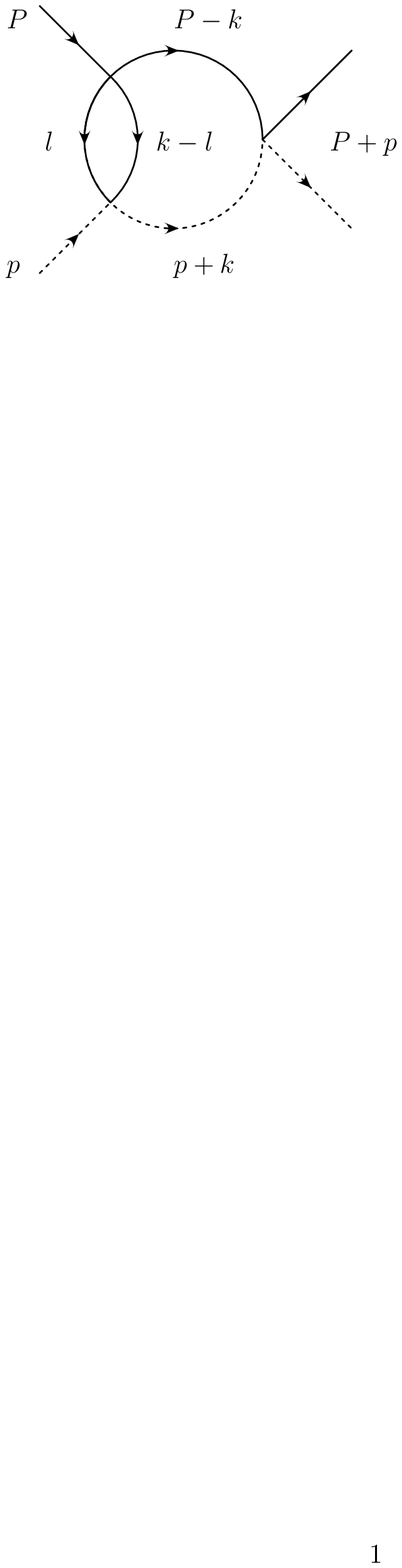,height=4cm,bbllx=120,bblly=645,bburx=300,bbury=778,clip}
\caption{Fish diagram with an internal kaon loop.}
\label{fig: KKloop}
\end{center}
\end{figure}
The fish graph shown in Fig. \ref{fig: KKloop} looks as follows
\begin{align}
F2 = \int_{k,l} \frac{1}{M^2-l^2-i\eps}\frac{1}{M^2-(k-l)^2-i\eps}\frac{1}{2Pk-k^2-i\eps}\frac{1}{-2pk-k^2-i\eps}
\fs
\end{align}
The leading contribution stems again from the h-h region, the s-h part starts contributing at order $m/M$, while the h-h', h-s and s-s contributions all vanish identically.

The expansion of the h-h region ($k\sim M$, $l \sim M$) takes the form
\begin{align}
F2^{\rm hh}& = \int_{k,l} \frac{1}{M^2-l^2-i\eps}\frac{1}{M^2-(k-l)^2-i\eps}\frac{1}{2Pk-k^2-i\eps}\frac{1}{-k^2-i\eps} 
\no
&\quadp\times
\left\{ 1+\frac{2pk}{-k^2-i\eps}+O\left(\frac{m^2}{M^2}\right)\right\} \non\\
&= F2^{\rm hh}_0+F2^{\rm hh}_1+O\left(\frac{m^2}{M^2}\right)
\fs
\end{align}
As with the s-s contributions to the pion loop diagram discussed above,  we may use integration by parts to reduce $F2^{\rm hh}$ to the sunset type diagram $S_{1,1,2}$ 
\begin{align}
F2^{\rm hh}_0 = \frac{1}{d-4}\left[2S_{1,1,2}\{M,M,M;M^2\}-I_2\{M\}I_{1,1}\{0,M;M^2\}\right]
\fs
\end{align}
At leading order $F2$ amounts to
\begin{align}
F2_0 &=  F2^{\rm hh}_0 
\\
& = \frac{\mubar^{-4\epsj}  }{(4\pi)^4}\left\{\frac{1}{2\epsj^2}+\frac{1}{2\epsj}\left[5-2\ln \frac{M^2}{\mu^2}\right]+\ln^2 \frac{M^2}{\mu^2}-5\ln \frac{M^2}{\mu^2}+\frac{19}{2}-\frac{7\pi^2}{12}\right\} +O(\epsj) 
\fs \non  
\end{align}
At order $m/M$, both the h-h and s-h regions ($k\sim m$, $l\sim M$) contribute to $F2$. The h-h part takes the form
\begin{align}
F2^{\rm hh}_1 =  -\frac{m}{M} \left[F2^{\rm hh}_0 - S_{1,1,2}\{M,M,0;0\} \right] 
\co
\end{align}
The result for the sunset integral $S_{1,1,2}\{M,M,0;0\}$ can be given in terms of Gamma functions and is listed in Appendix~\ref{app: Two loop Integrals}. The s-h contribution yields
\begin{align}
F2^{\rm sh}_1 &= \int_{k,l} \frac{1}{[M^2-l^2-i\eps]^2}\frac{1}{2Pk-i\eps} \frac{1}{-2p k - k^2-i\eps}\non\\
&= \frac{m}{M}\, I_2\{M\}I_{1\strichpunkt1}\{m;m^2,m^2\} 
\fs
\end{align}
In total, we get for $F2$ at next-to-leading order 
\begin{align}
F2_1&= F2_1^{\rm hh} + F2_1^{\rm sh} 
\\
&= \frac{m}{M}\frac{\mubar^{-4\epsj} }{(4\pi)^4}\left\{-\left[ \frac{1}{\epsj} - 2 \ln \frac{M^2}{\mu^2} \right] \ln \frac{m^2}{M^2}  + \frac{1}{2}\ln^2 \frac{m^2}{M^2} -2\ln \frac{m^2}{M^2}-7+\frac{2\pi^2}{3}\right\} +O(\epsj)
\fs \non 
\end{align}


\subsection{$\pi K$ loop}
\begin{figure}[t]
\begin{center}
\epsfig{figure=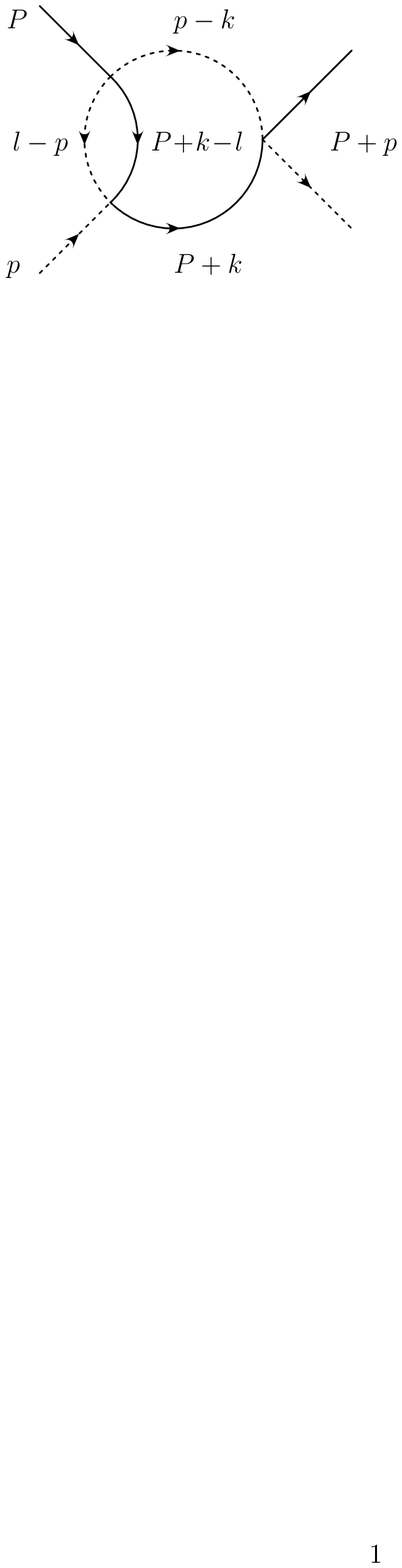,height=4cm,bbllx=120,bblly=645,bburx=300,bbury=778,clip}
\caption{Fish diagram with an internal $\pi K$ loop.}
\label{fig: piKloop}
\end{center}
\end{figure}
The fish diagram with an internal $\pi K$ loop, shown in Fig.~\ref{fig: piKloop},  is given by
\begin{align}
F3 = \int_{k,l} \frac{1}{ 2 p l -l^2-i\eps} \frac{1}{-2 P(k-l)-(k-l)^2 -i\eps} \frac{1}{2pk-k^2-i\eps} \frac{1}{-2Pk-k^2-i\eps}
\fs
\end{align}
The leading order contribution is generated by the h-h region, the s-h part contributes at order $m/M$, while the contributions from the h-s and s-s regions are of higher order in $m/M$ and the h-h' part vanishes identically.
 
The expansion of the h-h region ($k\sim M$, $l\sim M$) looks as follows
\begin{align}
F3^{\rm hh} &= 
 \int_{k,l} \frac{1}{ -l^2-i\eps} \frac{1}{-2 P(k-l)-(k-l)^2 -i\eps} \frac{1}{-k^2-i\eps} \frac{1}{-2Pk-k^2-i\eps}
 \no
 &\quadp \times \left\{ 1-\frac{2p k}{-k^2-i\eps}-\frac{2p l}{-l^2-i\eps} +O\left(\frac{m^2}{M^2}\right)\right\} \no
 &=F3^{\rm hh}_0+F3^{\rm hh}_1+O\left(\frac{m^2}{M^2}\right) \fs
 \end{align}
To evaluate the h-h part at leading order, we again make use of integration by parts, in this case by inserting $ 1 = \frac{1}{d}\frac{\partial}{\partial l_\mu} l_\mu$. Integrals of this type were also discussed in Refs.~\cite{Czarnecki:2000fv,Blokland:2001fn}. The result takes the form
\begin{align}
F3^{\rm hh}_0 = \frac{M^{2d-8}}{d-3}\left[ I_2\{1\}I_{1,1}\{0,1;1\}-I_{1,1}\{0,0;-1\}I_{2-\frac{d}{2},2}\{0,1;1\}\right] \co
 \end{align}
and $F3$ reads at order one
\begin{align}
F3_0 &= F3^{\rm hh}_0 \non\\
&= \frac{\mubar^{-4\epsj}}{(4\pi)^4}\left\{\frac{1}{2\epsj^2}+\frac{1}{2\epsj}\left[5-2\ln \frac{M^2}{\mu^2}\right]+\ln^2 \frac{M^2}{\mu^2}-5\ln\frac{M^2}{\mu^2}+\frac{19}{2}-\frac{\pi^2}{4}\right\} +O(\epsj)
\fs
 \end{align}
At order $m$ the function $F3$ receives contributions from the h-h 
\begin{align}
F3^{\rm hh}_1 =\frac{m}{M}\, M^{2d-8} \Big[ & -I_2\{1\}I_{1,1}\{0,1;1\} +I_{1,1}\{0,0;-1\}I_{2-\frac{d}{2},2}\{0,1;1\}
\no
&+2I_{1,2}\{0,0;-1\}I_{3-\frac{d}{2},1}\{0,1;1\} +S_{2,1,1}\{1,1,0;0\}\Big] \co 
 \end{align}
 and s-h regions ($k\sim m$, $l\sim M$)
 \begin{align}
 F3_1^{\rm sh} &= \int_{k,l} \frac{1}{ -l^2-i\eps} \frac{1}{2 Pl -l^2 -i\eps} \frac{1}{2pk-k^2-i\eps} \frac{1}{-2Pk-i\eps}  
 \no
 & = \frac{m}{M}\,I_{1,1}\{0,M;M^2\}I_{1\strichpunkt1}\{m;m^2,m^2\} \fs
 \end{align}
 The result for $F3^{\rm hh}_1$ can be obtained by inserting $  \frac{\partial}{\partial k_\mu}k_\mu =  \frac{\partial}{\partial l_\mu} (l_\mu-k_\mu) $ in $F3^{\rm hh}_0$. The net result for $F3_1$ amounts to
  \begin{align}
  \label{eq: F3}
F3_1&= F3_1^{\rm hh} + F3_1^{\rm sh} \\
&= \frac{m}{M}\frac{\mubar^{-4\epsj} }{(4\pi)^4}\left\{-\left[ \frac{1}{\epsj} - 2 \ln \frac{M^2}{\mu^2} \right] \ln \frac{m^2}{M^2} + \frac{1}{2}\ln^2 \frac{m^2}{M^2} -4\ln \frac{m^2}{M^2}-1- \frac{\pi^2}{3}\right\} +O(\epsj) \fs \non 
\end{align}

\section{Conclusions {and an application}}
\label{sec:conclusions}

As discussed before, the divergent part of the fish diagrams $F1$, $F2$ and $F3$ is known. When subtracting the poles in $d-4$ in a scale independent manner (cf. Appendix \ref{app: Two loop Integrals}), the results obtained in the previous section may be written in the compact form,
\begin{align}
\Fbar{1} & = \frac{1}{(4\pi)^4} \left\{ \frac{19}{2} + \frac{5 \pi^2}{12} + \frac{m}{M}\left[ \ln^2 \frac{m^2}{M^2} - 5 \ln \frac{m^2}{M^2} - 3 \pi^2  \right] \right\} + O\left(\frac{m^2}{M^2} \right) 
\co \no
\Fbar{2} & = \frac{1}{(4\pi)^4} \left\{ \frac{19}{2} - \frac{7 \pi^2}{12}+ \frac{m}{M}\left[ \frac{1}{2} \ln^2 \frac{m^2}{M^2}  - 2 \ln \frac{m^2}{M^2} -7 +\frac{2\pi^2}{3}  \right] \right\} + O\left(\frac{m^2}{M^2} \right) 
\co \no
\Fbar{3} & = \frac{1}{(4\pi)^4} \left\{ \frac{19}{2} - \frac{\pi^2}{4}+ \frac{m}{M}\left[ \frac{1}{2}\ln^2 \frac{m^2}{M^2}  - 4 \ln \frac{m^2}{M^2} -1 - \frac{\pi^2}{3}  \right] \right\} + O\left(\frac{m^2}{M^2} \right)  
\fs 
\end{align}
The equations imply that our calculation has reproduced the correct divergent pieces. We are not aware of an independent determination of the finite parts. We note that for $m/M= M_{\pi^+}/M_{K^+} \simeq 0.28 $ a suppression of the subleading terms is clearly observed, except in the case of $\Fbar{2}$ where the leading term happens to be small. 

These results are obtained by expanding the integrands for soft ($m$) and hard ($M$) loop momenta and subsequently summing the integrals of all contributions, to the desired order in $m/M$.      
The coefficients of the expansion are then given by homogeneous Feynman integrals. We demonstrate in some detail how those can be evaluated using integration by parts identities. The most difficult integral encountered is an equal mass on-shell sunset type graph which is is evaluated by standard methods in Appendix \ref{app: Two loop Integrals}. In addition to the two loop results we demonstrate the validity of the expansion by regions in the case of a one loop example to all orders in $m/M$.      

For the method  to work, it is necessary to route the hard external momentum through the heavy particle line. In the present case, the conservation of Strangeness guarantees that this can always be done. For the same reason the problem of $\pi K $ scattering in QCD can be attacked also in an effective theory where the kaons are treated as heavy from the start \cite{Roessl:1999iu}.\footnote{Similar are the cases of Heavy Quark Effective Theory \cite{HQET} or (Heavy) Baryon Chiral Perturbation Theory \cite{Gasser:1987rb,Jenkins:1990jv,Bernard:1992qa,Ecker:1993ft,Bernard:1995dp,Becher:1999he}.}  In fact, the method of the expansion by regions and the formulation in terms of an effective theory are very closely related \cite{Beneke:1997zp,Smirnov:2002pj}. 

The approach followed in Refs.~\cite{Schweizer:2005nn,Haefeli} and in the present work is a different one: Since the $\pi K$ scattering amplitude is already known to two loop order in SU(3) chiral perturbation theory \cite{Bijnens:2004bu} one would like to make use of those results to obtain the scattering lengths. The integrals discussed in the present work are believed to be the most intricate ones in what concerns the expansion in $M_\pi/ M_K$ so that we hope to have contributed a significant step towards this goal.   

The result for the $\pi K$ scattering amplitude \cite{Bijnens:2004bu} at threshold also involves diagrams with $\eta$ mesons running in the loops.\footnote{For the S-channel, there are six additional diagrams containing $\eta$ mesons.} Note that the mass of the $\eta$ is not independent but can be expressed in terms of the pion and kaon masses by virtue of the Gell-Mann-Okubo relation, 
\begin{align}
\label{GMO}
M_\eta^2 = \frac{1}{3}  (4 M_K^2 - M_\pi^2 )   + O(m_{\rm quark}^2) \fs
\end{align}
Also in this case one therefore deals with a problem with two mass scales and the expansion by regions proceeds exactly as in the examples considered above. It should furthermore be obvious that the applicability of the method is not limited to scattering problems but that it could as well be applied to a similar analysis of masses and decay constants or form factors in SU(3) chiral perturbation theory. 
{As a first application we have calculated the leading strange quark contributions to the pion mass
\begin{align}
\label{mpi2loops}
\frac{M_\pi^2}{(m_u+m_d)B_0} & = 1 + c_1 \frac{m_s B_0}{(4 \pi F_0)^2} + c_2 \left[ \frac{m_s B_0}{(4 \pi F_0)^2}\right]^2+ O(m_u,m_d,m_s^3)
\co \\
c_1 &= -16 (4 \pi)^2 (L_4^r-2 L_6^r) -  \frac{2}{9} \ln \frac{4 m_s B_0}{3 \mu^2} 
\co \no
c_2 & = c_2^{\rm tree} + c_2^{\rm loop} 
\co \no  
c_2^{\rm tree} &= 64 (4 \pi)^4  \left[ (-C_{16}^r +C_{20}^r +3 C_{21}^r) F_0^2 + 4 L_4^r (L_4^r  -2 L_6^r) \right] 
\co \no  
c_2^{\rm loop} &= \frac{11}{12} \ln^2 \frac{ m_s B_0 }{\mu^2}  - \left[\frac{32}{9} {\cal L}_1^r + \frac{380}{81}-\frac{2}{9} \ln \tfrac{4}{3} \right]   \ln \frac{ m_s B_0 }{\mu^2}
 + \frac{16}{9}  \left[ {\cal L}_2^r- 2 \ln \tfrac{4}{3} \, {\cal L}_3^r \right] 
 \co \no 
& \quadp + \frac{73}{16} - \frac{38}{81} \ln \tfrac{4}{3} + \frac{2}{9} \ln^2 \tfrac{4}{3}  -\frac{4 \sqrt{2}}{3} \, {\rm Im } \left[ {\rm Li_2 } \, \frac{1 + 2\sqrt{2} i }{3}\right] 
\co \non 
\end{align}
where ${\rm Li}_2$ is the dilogarithm, ${\rm Li}_2(z) = \sum_{n=1}^\infty \frac{z^n}{n^2} $~\cite{Lewin}. We have made use of the representation for the pion mass of Ref.~\cite{Amoros:1999dp}, for the definitions of the coupling constants see Refs.~\cite{Gasser:1984gg,Bijnens:1999sh} (Note that our $C_i^r$ carry mass dimension $-2$). 
We have also introduced the following auxiliary symbols for combinations of the $L_i$,
\begin{align}
{\cal L}_1^r & ={(4\pi)^2} \left[ 26 L_1^r +\frac{13}{2} L_2^r + \frac{61}{8} L_3- 29 L_4^r -\frac{13}{2} L_5^r +30 L_6^r -6 L_7 +11 L_8^r \right]
\co \\ 
{\cal L}_2^r & ={(4\pi)^2} \left[ \frac{13}{2} L_2^r + \frac{43}{24} L_3 +2 L_4^r + \frac{4}{3} L_5^r -4( L_6^r + L_7 +L_8^r)  \right]
\co \no 
{\cal L}_3^r & ={(4\pi)^2} \left[ 8 L_1^r + 2(L_2^r +  L_3) - 11 L_4^r -2  L_5^r+12 L_6^r -6 L_7 +2  L_8^r \right]
\fs \non 
\end{align}
Taking the scale dependence of the chiral coupling constants into account~\cite{Gasser:1984gg,Bijnens:1999hw}, one verifies that the coefficients $c_1$ and $c_2$ are separately independent of the chiral renormalization scale $ \mu $ (Note that the product $m_s B_0$ is QCD scale independent).  
The relevance of the formula Eq.~(\ref{mpi2loops}) lies in the fact that it fully specifies the dependence of the SU(2) coupling constant $B$~\cite{Gasser:1983yg} on the strange quark mass $m_s $ at two loops, 
\begin{align}
B = B_0 \left\{ 1 + c_1 \frac{m_s B_0}{(4 \pi  F_0)^2} + c_2 \left[ \frac{m_s B_0}{(4 \pi F_0)^2}\right]^2  \right\} + O(m_s^3)
\fs  
\end{align}
The term $ c_1$ agrees with the result of Ref.~\cite{Gasser:1984gg}. For $\sqrt{m_s B_0} = 484 \,{\rm MeV} $ and $\mu = 770 \,{\rm MeV}$ the logarithm in $c_1$ amounts to about $0.14$. Neglecting the terms proportional to the $L_i$ and $C_i$ in $c_2$ one finds a value of about $7.6$. The bulk of this contribution stems from the term 
$ - {380}/{81} \ln ({ m_s B_0 }/{\mu^2}) \simeq 4.4$ while the contribution from the last line amounts to about $2.6$ ($ {\rm Im } [ {\rm Li_2 } \, \frac{1 + 2\sqrt{2} i }{3} ] \simeq 1.001 $). With those values and $F_0 = 92 \,{\rm MeV} $ as estimates, the expansion takes the rather peculiar form $ B/B_0 = 1+ 0.02 + 0.23 + \ldots$. 
The presumably important impact of the contributions from the coupling constants $L_i$ and $C_i$ remains to be investigated.}\footnote{Note that the relevant combination $C_{16}^r -C_{20}^r -3 C_{21}^r $ is Zweig rule violating. In the estimate of Ref.~\cite{Cirigliano:2006hb} it therefore fails to receive a nonvanishing contribution.} We believe that further such studies could provide substantial insight on the convergence of the chiral perturbation series.  

\begin{acknowledgments}
We are indebted to J.~Gasser for the motivation to undertake the present work. 
We are are also grateful to T.~Becher, J.~Bijnens, G.~Ecker, J.~Gasser, D.~Greynat, E.~de Rafael and P.~Talavera for suggestions and comments on the manuscript. {We thank C.~Haefeli  and M.~Schmid for useful correspondence in relation to the pion mass formula and M.~Schmid for providing us with his results prior to publication.}   
R.K. was supported by the Swiss National Science Foundation and J.S. was supported by EC-Contract HPRN-CT2002-00311 (EURIDICE).
\end{acknowledgments}

\appendix

\section{Table of one loop integrals}
\label{app: Table of one loop Integrals}
The present Appendix lists results for a number of Feynman integrals that appear in the main text.  
Since the one loop integrals occur with various configurations of the indices, it is convenient to list them here in general form, leaving also the dimension $d$ arbitrary. In cases of a singular configuration of indices, the relevant result is obtained via a limiting procedure. For a more complete list of this kind, see e.g. Refs.~\cite{Smirnov:2002pj,Smirnov:2004ym}. Throughout, we use the following shorthand notation
\begin{align}
\int_{k} \equiv \frac{1}{i} \int  \frac{d^d k}{(2 \pi)^d} 
\fs
\end{align}
We encounter various integrals of type one point or two point function,
\begin{align}
\label{Ialpha}
I_\alpha\{M \}  & =  \int_k  \frac{1}{[M^2-k^2-i \eps]^\alpha} = \frac{1}{(4 \pi)^\frac{d}{2}}\frac{\Gamma(\alpha-\frac{d}{2})}{\Gamma(\alpha)}[M^2- i\eps]^{\frac{d}{2}-\alpha} 
\co \\
I_{\alpha,\beta}\{M_1,M_2;p^2 \}    & =  \int_k  \frac{1}{[M_1^2-k^2-i \eps]^\alpha} \frac{1}{[M_2^2-(p-k)^2-i \eps]^\beta} \fs 
\end{align}
In this notation the $J $ integral at threshold is given by
\begin{align}
J_{\rm thr}(m,M)    = I_{1,1} \{m,M;(m+M)^2\}  =  \frac{1}{( 4\pi)^\frac{d}{2}} \frac{\Gamma(2-\frac{d}{2})}{d-3} \frac{m^{d-3}+ M^{d-3}}{m+M}  \fs
\end{align}
For vanishing masses, $M_1 = M_2 =0 $, or in the case of the integral at threshold with one vanishing mass, $M_1 = 0$, $p^2 =M_2^2 = M^2$, one has
\begin{align}
I_{\alpha,\beta} \{ 0,0; p^2\}  & = \frac{1}{(4 \pi)^\frac{d}{2}}
\frac{\Gamma(\alpha+\beta-\frac{d}{2})\Gamma(\frac{d}{2}-\alpha)\Gamma(\frac{d}{2}-\beta)}{\Gamma(\alpha)\Gamma(\beta)\Gamma(d-\alpha-\beta)} [-p^2 - i\eps]^{\frac{d}{2}-\alpha-\beta} 
\co \\
I_{\alpha,\beta}\{ 0,M;M^2 \}  & =  \frac{1}{(4 \pi)^\frac{d}{2}}\frac{\Gamma(\alpha+\beta-\frac{d}{2})\Gamma(d-2 \alpha- \beta)}{\Gamma(\beta)\Gamma(d-\alpha-\beta)}\  [M^2-i\eps]^{\frac{d}{2}-\alpha-\beta} \fs
\end{align}
A different class of integrals involves a denominator which is linear in the loop momentum: 
\begin{align}
I_{\alpha\strichpunkt\beta}\{M; p \,p',p'^2 \}   & =  \int_k  \frac{1}{[M^2-(p-k)^2-i \eps]^\alpha}\frac{1}{[-2 p' k-i \eps]^\beta} \fs 
\end{align}
For $p= 0$ one finds
\begin{align}
\label{IaIbp=0}
I_{\alpha\strichpunkt\beta}\{ M ;0 ,p'^2 \} & =  \frac{1}{(4 \pi)^\frac{d}{2}}\frac{\Gamma(\alpha+\frac{\beta}{2}-\frac{d}{2})\Gamma(\frac{\beta}{2})}{2 \Gamma(\alpha)\Gamma(\beta)}\,  [M^2-i\eps]^{\frac{d}{2}-\alpha-\frac{\beta}{2}} [p'^2-i \eps ]^{-\frac{\beta}{2}}
\co 
\end{align}
while for $p^2 = M^2$, $ p' = \pm p $ and integer $\beta$ one has 
\begin{align}
I_{\alpha\strichpunkt\beta}\{M; \pm M^2, M^2\} & =  \frac{(\mp1)^\beta}{(4 \pi)^\frac{d}{2}}\frac{\Gamma(\alpha+\beta-\frac{d}{2})\Gamma(2 \alpha + \beta -d)}{\Gamma(\alpha)\Gamma(2 \alpha+ 2 \beta-d)}\, [M^2-i\eps]^{\frac{d}{2}-\alpha-\beta} 
\fs
\end{align} 


\section{Two loop integrals} 
\label{app: Two loop Integrals}

In this appendix we collect various results for the sunset and fish integrals. We use the shorthand notation
\begin{align}
\int_{k,l} \equiv \frac{1}{i^2} \int  \frac{d^d k}{(2 \pi)^d}  \int  \frac{d^d l}{(2 \pi)^d} \fs
\end{align}

\subsection{Sunset type integrals}

\begin{figure}[t]
\begin{center}
\epsfig{figure=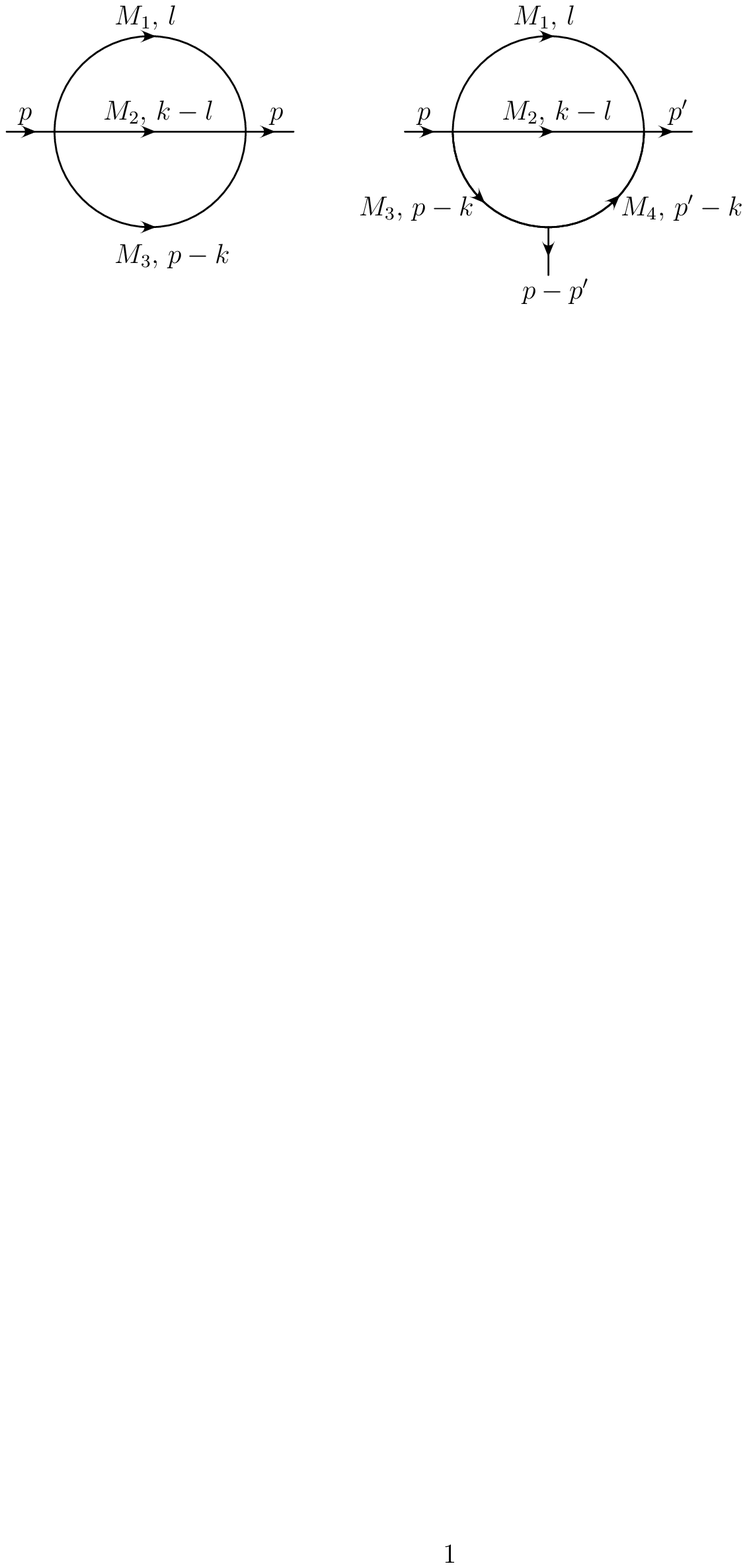,height=4.5cm,bbllx=90,bblly=630,bburx=418,bbury=778,clip}
\caption{The sunset and fish topologies.}
\label{fig: sunset}
\end{center}
\end{figure}

The general integral of type sunset,\footnote{The integral also goes by the names of `sunrise' or `setting sun diagram'.} shown in Fig.~\ref{fig: sunset}, takes the form
\begin{align}
& \hspace*{-4em}  S_{\alpha,\beta,\gamma} \{M_1 , M_2, M_3   ; p^2 \}  =   
\no& 
\int_{k,l}   \frac{1}{[M_1^2-l^2-i \eps]^\alpha} \frac{1}{[M_2^2-(k-l)^2-i \eps]^\beta} \frac{1}{[M_3^2-(p-k)^2-i \eps]^\gamma} 
\fs 
\label{eq: sunset def}
\end{align}

The following Feynman parameter representation for the integral is obtained by first performing the integration over the loop momentum $l$ in Eq.~(\ref{eq: sunset def}) and subsequently combining the two remaining denominators~\cite{Ramond},
\begin{align}
\label{eq: FeynPSunset}
& S_{\alpha,\beta,\gamma} \{M_1, M_2 , M_3  ; p^2 \} =   \frac{1}{(4 \pi)^{d}} \frac{\Gamma(\alpha+\beta+\gamma-d)}{\Gamma(\alpha)\Gamma(\beta)\Gamma(\gamma)} \int_0^1 \! dx \, x^{\frac{d}{2}-\beta-1} (1-x)^{\frac{d}{2}-\alpha-1}     
\\
& \hspace{4em} \times     \int_0^1 \! dy \, y^{\alpha+\beta- \frac{d}{2}-1} (1-y)^{\gamma-1}  [y  M_x^2+ (1-y)M_3^2  -y(1-y) p^2 - i \eps ]^{d-\alpha-\beta-\gamma} 
\fs \non
\end{align}
The combination $M_x$ depends on $M_1$, $M_2$ as well as on the Feynman parameter $x$
\begin{align}
M_x^2 & =  \frac{M_1^2}{1-x}+\frac{M_2^2}{x}
\fs 
\end{align}

In the following special cases 
the sunset integral may be expressed in terms of Gamma functions, 
\begin{align}
S_{\alpha ,\beta,\gamma} \{M,M, 0 ; 0 \}  & = \frac{1}{(4\pi)^d} \frac{\Gamma(\alpha+\beta+\gamma-d)}{\Gamma(\alpha+\beta+2\gamma-d)} 
\frac{\Gamma(\alpha+\gamma-\frac{d}{2})\Gamma(\beta+\gamma-\frac{d}{2})\Gamma(\frac{d}{2}-\gamma)}{\Gamma(\alpha)\Gamma(\beta)\Gamma(\frac{d}{2})}  
\no  & \quadp
\times 
[M^2-i\eps]^{d-\alpha-\beta-\gamma}  
\co \no 
S_{\alpha ,\beta,\gamma} \{0,0, M ; M^2 \}  & = I_{\alpha,\beta} \{ 0,0;-1\} \, I_{\alpha+\beta-\frac{d}{2},\gamma} \{ 0,1;1\}  \, [M^2-i\eps]^{d-\alpha-\beta-\gamma} 
\co 
\end{align}
where the last result follows directly from Eq.~(\ref{eq: sunset def}).  

The evaluation of the diagrams in the main text
does also require the result for the equal mass on-shell sunset integral with one denominator raised to the second power,
viz. $S_{1,1,2} \{M,M,M ; M^2 \}$. Furthermore, its contribution comes with a prefactor $1/(d-4)$ so that we need the expression
including contributions of $O(d-4)$. While the result in question is implicit in Ref. \cite{Laporta:2004rb}, we 
pursued a more direct approach following Ref.~\cite{Ramond}: In the Feynman parameter integral representation write ($\epsj = 2-d/2$)  
\begin{align}
y^{-1+\epsj} = \frac{1}{\epsj}\frac{d}{dy}\,  y^\epsj 
\end{align}
and integrate by parts to obtain
\begin{align}
S_{1,1,2}\{M,M,M;M^2\} &= 
[M^2-i \eps]^{-2\epsj} \frac{\Gamma(2\epsj)}{\epsj (4\pi)^{4-2\epsj}}
\int_0^1 \!dx \,[x(1-x)]^{-\epsj}\int_0^1 \!dy\, y^\epsj 
\\
&\hspace{-5.5em} \times
\left\{1+2\epsj(1-y)\frac{d}{dy}\ln \left[\frac{y}{x(1-x)}+(1-y)^2\right]\right\}
 \left\{\frac{y}{x(1-x)}+(1-y)^2\right\}^{-2\epsj} 
 \fs \non
\end{align}
Now the parametric integrals can be done by expanding the integrand around $\epsj = 0$. We obtain
\begin{align}
\label{eq: sunset}
S_{1,1,2} \{M,M,M ; M^2 \}  & =
\frac{\Gamma(\epsj)^2}{2(4\pi)^{4-2 \epsj}} \{ 1 + \epsj- \epsj^2 - \frac{33-4\pi^2}{3}\, \epsj^3 \} [M^2-i\eps]^{-2 \epsj}   
+ O(\epsj^2)
\fs 
\end{align}
We observe that this result is of equal simplicity as the on-shell sunset itself \cite{Argeri:2002wz, Broadhurst:1991fi, Gasser:1998qt, Laporta:2004rb}
\begin{align}
S_{1,1,1} \{M,M,M ; M^2 \}  & =
\frac{\Gamma(\epsj)^2}{2(4\pi)^{4-2 \epsj}} \{-3-\frac{17}{2} \, \epsj-\frac{59}{4}\, \epsj^2 - \frac{195 + 64 \pi^2}{24 }\,  \epsj^3 \} [M^2-i\eps]^{1-2\epsj}  
\no
& \quadp + O(\epsj^2)
\fs  
\end{align}

Finally, we also encounter a sunset diagram with one denominator which is linear in the loop momentum
\begin{align}
S_{\alpha,\beta \strichpunkt \gamma} \{M_1 , M_2; p p '   , p'^2 \}  =   
\int_{k,l}   \frac{1}{[M_1^2-(p-l)^2-i \eps]^\alpha} \frac{1}{[M_2^2-(l-k)^2-i \eps]^\beta} \frac{1}{[-2  p' k-i \eps]^\gamma} 
\fs 
\label{eq: sunset type 2 def}
\end{align}
For $M_1 =M_2 =M $ and $ p = 0 $ we have 
\begin{align}
S_{\alpha,\beta\strichpunkt \gamma} \{ M, M ; 0 , p'^2 \}  & =  \frac{1}{(4\pi)^d} 
\frac{\Gamma(\alpha+\beta+\frac{\gamma}{2}-d)}
{\Gamma(\alpha+\beta+ \gamma -d) } 
\frac{\Gamma(\alpha+\frac{\gamma}{2}-\frac{d}{2})\Gamma(\alpha+\frac{\gamma}{2}-\frac{d}{2})\Gamma(\frac{\gamma}{2})}{2 \Gamma(\alpha) \Gamma(\beta)\Gamma(\gamma)}  
\no  & \quadp
\times 
[M^2-i\eps]^{d-\alpha-\beta-\frac{\gamma}{2}}   [p'^2-i\eps]^{-\frac{\gamma}{2}}   
\co
\end{align}
which for $\alpha = \beta =1 $ agrees with the result given in Ref.~\cite{Czarnecki:1996nr}. To obtain it, combine the first two denominators and perform the $ l $ integration. After this, the formula Eq.~(\ref{IaIbp=0}) can be applied.  

\subsection{Fish type integrals}

The general integral of type fish, shown in Fig.~\ref{fig: sunset}, is defined by
\begin{align}
\label{eq: fish def}
& F_{\alpha,\beta,\gamma,\delta} \{M_1 , M_2,M_3 ,M_4  ; p^2 ,p'^2,(p-p')^2 \}  = 
\\
& \int_{k,l}   \frac{1}{[M_1^2-l^2-i \eps]^\alpha} \frac{1}{[M_2^2-(k-l)^2-i \eps]^\beta} \frac{1}{[M_3^2-(p-k)^2-i \eps]^\gamma}  \frac{1}{[M_4^2-(p'-k)^2-i \eps]^\delta} 
\fs \non
\end{align}
In this notation the integrals  $F1$, $F2$ and $F3$ discussed in Sec.~\ref{sec:fish} are given by
\begin{align}
F1 & = F_{1,1,1,1} \{m ,m,M ,m ; M^2  ,m^2,(m+M)^2 \} 
\co \no 
F2 & = F_{1,1,1,1} \{M,M,M ,m ; M^2  ,m^2,(m+M)^2 \} 
\co \no
F3 & = F_{1,1,1,1} \{m,M ,M ,m ; m^2  ,M^2,(m+M)^2 \} 
\fs 
\end{align}
By combining the third and fourth denominators the diagram is lead back to the sunset \cite{Ghinculov:1994sd}, 
\begin{align}
&\hspace{-3em} F_{\alpha,\beta,\gamma,\delta} \{M_1, M_2 , M_3 ,M_4 ; p^2 , p'^2 ,(p-p')^2\}  =
\\
& \hspace{6em}   \frac{\Gamma(\gamma+\delta)}{\Gamma(\gamma) \Gamma(\delta)} \int_0^1 \! dz \,  z^{\gamma-1} (1-z)^{\delta-1} S_{\alpha,\beta,\gamma+\delta} \{M_1, M_2 , M_z ; p_z^2 \}  
\co \non
\end{align}
with
\begin{align}
M_z^2 & = z M_3^2 + (1-z)M_4^2 - z(1-z) (p-p')^2 - i \eps
\co  \quad 
p_z  = z p+(1-z)p' 
\fs 
\end{align}
This representation is useful e.g., for the calculation of the (nonlocal) divergent part of the fish diagram with unit  exponents, 
\begin{align}
\label{eq: fish div}
& \hspace{-2em}F_{1,1,1,1} \{M_1 , M_2,M_3 ,M_4  ; p^2 ,p'^2,(p-p')^2 \} = 
\\
& M_3^{d-4} \frac{\Gamma(2-\frac{d}{2})}{(4\pi)^\frac{d}{2}} \, I_{1,1} \{ M_3,M_4; (p-p')^2\}
+  \frac{M_3^{2d-8}}{d-6} \frac{\Gamma(2-\frac{d}{2})^2}{(4\pi)^d}  +O((d-4)^0)
\co \non
\end{align}
which, in the equal mass case, is seen to agree with the findings of Ref.~\cite{Gasser:1998qt}. 
The formula shows that the divergent part is in fact independent of the mass content of the subgraph. In particular, this implies that the divergent pieces of the integrals $F1$, $F2$ and $F3$ discussed in Sec.~\ref{sec:fish} are equal.
We can therefore define a quantity 
$\Fdiv$, 
\begin{align} 
\Fdiv =   \frac{\mubar^{2d-8}}{(4\pi)^4}  & \left\{ \frac{2}{(d-4)^2}- \frac{1}{d-4}\left[ 5-2 \ln \frac{M^2}{\mu^2}\right]+\ln^2 \frac{M^2}{\mu^2}-5\ln\frac{M^2}{\mu^2}  \right. 
\no
& \left. \quad + \frac{2 m}{m+M} \left[ \frac{1}{d-4}+\ln \frac{M^2}{\mu^2} \right] \ln \frac{m^2}{M^2}   \right\} \co
\end{align}
$ d \Fdiv /d\mu = O(d-4)$, such that the limiting expressions
\begin{align}
\Fbar{n} = \lim_{d \to 4} ( Fn -\Fdiv ) \co \quad n=1,2,3 \co 
\end{align} 
are both finite and scale independent. 


\end{document}